\documentclass[amsmath,amssymb,amsfonts,aps,pre,preprint,superscriptaddress,bibnotes,showpacs,showkeys,longbibliography]{revtex4-1}
\usepackage{epsfig} 
\usepackage[english]{babel}
\usepackage{physics}
\usepackage{color}
\usepackage{subcaption}
\usepackage{hyperref}
\newcommand*{\logten}{\mathop{\log_{10}}}

\begin{document}
\title{Analytic iteration procedure for solitons and traveling wavefronts with sources}

\author{Jonas Berx}
\affiliation{Institute for  Theoretical Physics, KU Leuven, B-3001 Leuven, Belgium}

\author{Joseph O. Indekeu}
\affiliation{Institute for  Theoretical Physics, KU Leuven, B-3001 Leuven, Belgium}

\date{July 26, 2019}

\begin{abstract}
A method is presented for calculating solutions to differential equations analytically for a variety of problems in physics.
An iteration procedure based on the recently proposed BLUES (Beyond Linear Use of Equation Superposition) function method is shown to converge for nonlinear ordinary differential equations. Case studies are presented for solitary wave solutions of the Camassa-Holm equation and for traveling wavefront solutions of the Burgers equation, with source terms. The convergence of the analytical approximations towards the numerically exact solution is exponentially rapid. In practice, the zeroth-order approximation (a simple convolution) is already useful and the first-order approximation is already accurate while still easy to calculate. The type of nonlinearity can be chosen rather freely, which makes the method generally applicable.
\end{abstract}

\maketitle

Solving differential equations (DEs) is of general interest in physics, since they are the language in which many physical laws are formulated. Standard summation or integration methods for linear problems, such as the Fourier transform and the Green function method \cite{Green}, are not indicated when nonlinearity is present. For nonlinear problems \cite{NLDE} numerical computation is often the only way to go and although this is satisfactory from the point of view of precision, the  physical insight and control gained by having explicit analytical forms at one's disposal, is missing. For example, the way parameters in a DE affect its solution is conspicuous in an analytical expression. Analytic methods for solving nonlinear DEs are sparse. The method we propose is complementary to the Simplest Equation method \cite{Vitanov}. It is also complementary to the soliton perturbation approach \cite{SP,optical} and to the Homotopy Analysis Method \cite{HA}. In contrast with these methods, our approach appears easier to apply without giving in on accuracy. 

In this Letter we pose and answer, from a physicist's point of view, the following fundamental questions. i) Can methods for solving a linear ordinary differential equation (ODE) serve as a basis for a useful iteration procedure for solving a nonlinear ODE? ii) Can one obtain explicit analytical solutions that are simple and accurate in a nonlinear physical problem? iii) For which types of nonlinearity can one hope that the strategy might work? We provide positive answers to the first two questions by developing the recently proposed Beyond-Linear-Use-of-Equation-Superposition (BLUES) function method \cite{BLUES} into a useful iteration procedure with exponentially rapid convergence of the successive analytical approximations. To answer the third question we start from a linear ODE for which the Green function is known, and show that a convergent iteration can be obtained when a nonlinear term is added.  The type of nonlinearity can be chosen rather freely. This demonstrates the general applicability of the method, beyond the range of applications envisaged in \cite{BLUES}. Possible extensions of the method to nonlinear {\em partial} differential equations are beyond the scope of the present work.

We briefly recall the method proposed in \cite{BLUES}. Let ${\cal N}_z$ be a nonlinear differential operator and suppose a piecewise analytic function $B(z)$ is known that solves (under suitable boundary conditions)
\begin{equation}\label{conceptN}
{\cal N}_z B(z) = \delta (z),
\end{equation}
where the Dirac delta function has been added as a source term. This term cancels a possible discontinuity of the highest
derivative of $B$ in $z=0$. Examples have been given in  \cite{prelim,BLUES} in the form of traveling wave solutions of reaction-diffusion-convection DEs related to the Fisher equation of population biology \cite{Murraytextbook}. 

Next, one looks for a related linear operator ${\cal L}_z$ so that $B(z)$ also solves
\begin{equation}\label{conceptL}
{\cal L}_z B(z) = \delta (z),
\end{equation}
which makes $B$ the Green function for ${\cal L}_z$. The proposition in \cite{BLUES} is to derive the linear operator from the nonlinear one using linearization about $z=0$. While this can provide the desired linear DE with the property \eqref{conceptL} in some cases, in other cases the desired linear DE can be found heuristically and it is not necessary to linearize the original DE, as we will demonstrate in an example.

Now we ask whether there is a systematic way to solve \eqref{conceptN} for an arbitrary source $\psi (z)$, knowing that the convolution product of $B$ and $\psi$ solves the linear DE with source $\psi$,
\begin{equation}\label{conceptLs}
{\cal L}_z (B\ast\psi (z)) = \psi (z)
\end{equation}
The proposal of \cite{BLUES} is to make use of the nonlinear differential operator ${\cal R}_z \equiv {\cal L}_z-{\cal N}_z$, called residual, and to calculate the solution to the nonlinear problem in the form $B\ast\phi $, so that
\begin{equation}\label{conceptNs}
{\cal N}_z (B\ast\phi (z)) = \psi (z)
\end{equation}
where $\phi$ can be calculated making use of the identity
\begin{equation}\label{iterationstart}
\phi (z)= \psi (z) + {\cal R}_z (B\ast\phi (z)),
\end{equation}
which can be iterated. 

To zeroth order,
\begin{equation}\label{zero}
\phi ^{(0)}(z)= \psi (z) 
\end{equation}
and the $n$th-order approximation ($n\geq 1$) is found through \cite{error}
\begin{equation}\label{higher}
\phi ^{(n)}(z)= \psi (z) + {\cal R}_z (B\ast\phi ^{(n-1)}(z))
\end{equation}
Note that the zeroth-order solution to \eqref{conceptNs} is simply the convolution $B\ast\psi$. This amounts to using superposition beyond the domain of the linear DE, whence the name BLUES function for $B(z)$. 
Testing the iteration procedure is the task to which we now turn.

Our first example is in fluid mechanics. We study the propagation of waves of height $u(x,t)$ on shallow water, as described by a variety of DEs among which the Korteweg-de Vries equation is the best known. In this context a source term represents an external force and DEs of this type are under active study \cite{KdVsource}. Here we focus on a related DE, the (dimensionless) 
Camassa-Holm equation \cite{camassa-holm1993} with dispersion parameter $\kappa \geq0$,
\begin{equation}\label{eq:camassa-holm}
 u_t + 2\kappa u_x - u_{xxt} + 3u u_x = 2u_xu_{xx} + uu_{xxx},
\end{equation}
where $u_t = \partial u/\partial t$ etc. For $\kappa>0$ smooth soliton solutions exist. These are traveling pulses that depend on the co-moving coordinate $z = x-ct$, with $c$ the velocity. For vanishing dispersion ($\kappa\rightarrow0$) the solitons become piecewise analytic with a jump in slope in their peak (``peakons"), which we place at $z=0$. The dispersionless nonlinear ODE, in the co-moving frame, can be integrated once and one obtains $ -c (U - U_{zz}) + \frac{3}{2}U^2 - \frac{1}{2}U_z^2 - UU_{zz} = 0$, with $U(z) \equiv u(x,t)$ and we have used the boundary conditions $U(|z|\rightarrow\infty) = 0$. 

In order for a peakon solution of arbitrary amplitude to be exact in every point including $z=0$, we add a co-moving Dirac delta function source of strength $s$ to cancel the jump in $U_z$, which leads to the nonlinear ODE 
\begin{equation}\label{eq:camassa-holm_integrated_dirac}
\mathcal{N}_zU \equiv \frac{1}{s}\left(-c (U - U_{zz}) + \frac{3}{2}U^2 - \frac{1}{2}U_z^2 - UU_{zz}\right) = \delta(z)
\end{equation}
This is solved exactly by the peakon solution 
\begin{equation}\label{eq:BLUES_peakon}
U(z) = A\mathrm{e}^{-|z|} \equiv B(z),
\end{equation}
provided the strength $s$ of the source, the wave speed $c$ and the peakon amplitude $A$ satisfy
$s = 2A(A -c)$. Note that for $s = 0$ we retrieve the peakon  with  $A=c$ of the dispersionless Camassa-Holm equation without source \cite{camassa-holm1993}. We now calculate analytically a solution for an arbitrary co-moving source $\psi(z)$. Following \cite{BLUES} we recall three lines of motivation for this source: a) to smoothen the singularity of the peakon, b) to include the effect of a co-moving agent or substance, or c) to add a co-moving external field or probe.

We now look heuristically for a related linear ODE for which the peakon is the Green function. It can be found by neglecting the difference $(U^2 - U_{z}^2)/2$ in \eqref{eq:camassa-holm_integrated_dirac} and linearizing the remaining terms about $z=0$ to zeroth order in $z$, setting $U(z) = A$, but keeping the derivatives. Indeed, the linear ODE obtained in this {\em ad hoc} way,
\begin{equation}\label{eq:camassa-holm_linearized}
\mathcal{L}_zU \equiv \frac{1}{s}\left(A-c\right) \left(U -U_{zz}\right) = \delta(z),
\end{equation}
with the same boundary conditions $U(|z|\rightarrow\infty) = 0$, is again solved by \eqref{eq:BLUES_peakon}.  Since the peakon solves the nonlinear DE with a Dirac delta source and is also the Green function for the related linear DE, it is a  BLUES function according to the definition proposed in \cite{BLUES}. 

We now consider the DE with an arbitrary source $\psi$ localised about $z=0$, which for convenience we normalize to unity,
$    \int\,\psi(z)\mathrm{d}z = 1$. We can take, e.g., the exponential corner source (cf. \cite{BLUES}),
\begin{equation}\label{eq:source}
    \psi(z) = \frac{\mathrm{e}^{-|z|/K}}{2K},
\end{equation}
which is itself a peakon with tunable decay length, tending
to a delta function in the limit $K\rightarrow0$. The zeroth-order solution is 
\begin{equation}\label{eq:zeroth_order}
    \begin{split}
       (B\ast\psi)(z) = \frac{A}{K^2 -1}\left(K \mathrm{e}^{-|z|/K} - \mathrm{e}^{-|z|}\right)
    \end{split}
\end{equation}
and the first-order solution takes the form 
\begin{equation}\label{eq:first_correction}
    \begin{split}
           (B\ast\phi^{(1)})(z) = (B\ast\psi)(z)  -\frac{2A^3K}{s(K^2-1)}\left[\frac{3\left(K\mathrm{e}^{-2|z|/K} -2\mathrm{e}^{-|z|}\right)}{2(K^2-4)}\right. +\left.\frac{\mathrm{e}^{-|z|(1+1/K)}}{K+1} - \mathrm{e}^{-|z|/K}\right]
    \end{split}
\end{equation}
Higher-order approximations can be calculated using \eqref{higher}. 

\begin{figure}[!ht]
\begin{subfigure}{0.60\linewidth}
    \centering
    \includegraphics[width = \linewidth]{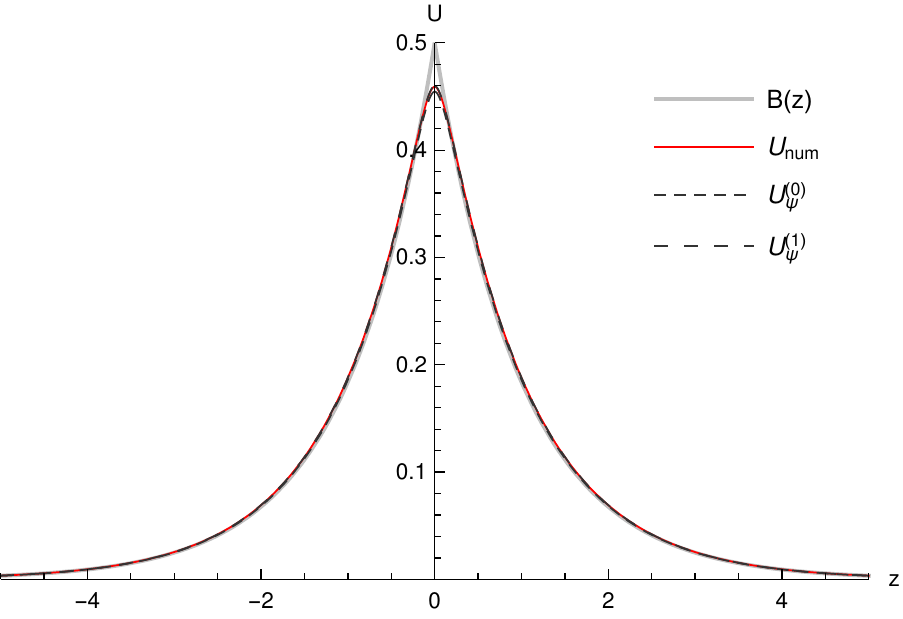}
    \caption{}
    \label{fig:ch_k05}
\end{subfigure}
\begin{subfigure}{0.60\linewidth}
    \centering
    \includegraphics[width = \linewidth]{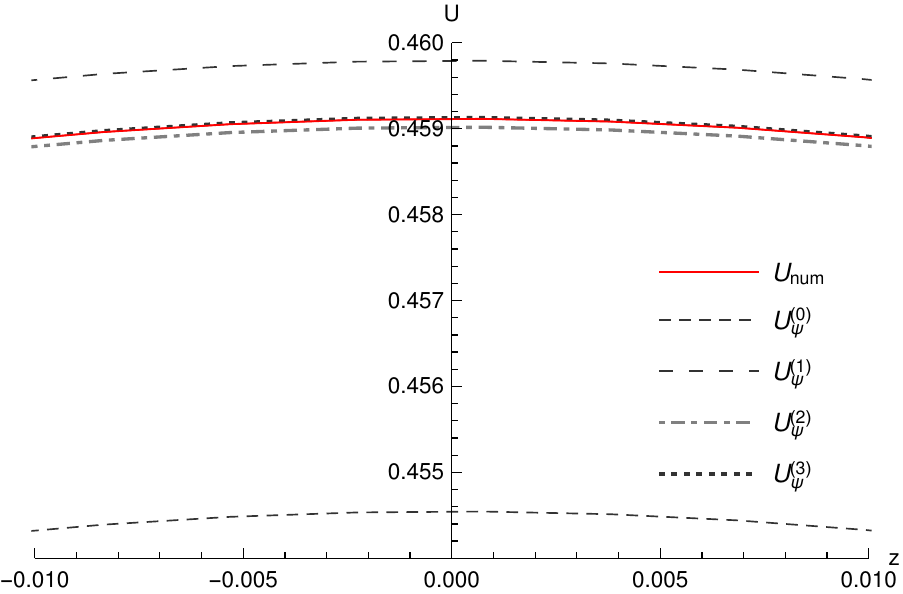}
    \caption{}
    \label{fig:ch_zoom_k05}
\end{subfigure}
\caption{ \textbf{(a)} Soliton solution to the nonlinear Camassa-Holm DE \eqref{eq:camassa-holm_integrated_dirac} with the exponential corner source \eqref{eq:source}. The numerical solution $U_{\rm num}$ (red, full line), the zeroth-order approximation $U_\psi^{(0)}$ (black, dashed line) and the first-order one $U_\psi^{(1)}$ (black, wider spaced dashed line) are compared. The BLUES function (gray) is also shown. On this scale $U_\psi^{(1)}$ is on top of the numerical solution. Parameter values are $c = -3/2$, $A = 1/2$, and $K = 1/2$. \textbf{(b)} A zoomed-in view around the maximum. The approximations are shown up to and including 3rd order. On this scale $U_\psi^{(3)}$ is on top of the numerical solution. }
\end{figure}

We now compare the approximate analytical solutions with the numerically exact solution of the nonlinear DE  \eqref{eq:camassa-holm_integrated_dirac} with source \eqref{eq:source}. The zeroth-order approximation (simple convolution) follows the numerical solution well and higher-order approximations systematically improve upon this. In Fig.\ref{fig:ch_k05}  the BLUES function, and the zeroth-order and first-order approximations are shown, together with the numerical solution, for $K=1/2$. In Fig.\ref{fig:ch_zoom_k05} a zoom is presented near the peak at $z=0$ and solutions are shown up till 3rd order. Even for this relatively broad source the first-order approximation \eqref{eq:first_correction} is already accurate. 

\begin{figure}[!ht]
    \centering
    \includegraphics[width = 0.60\linewidth]{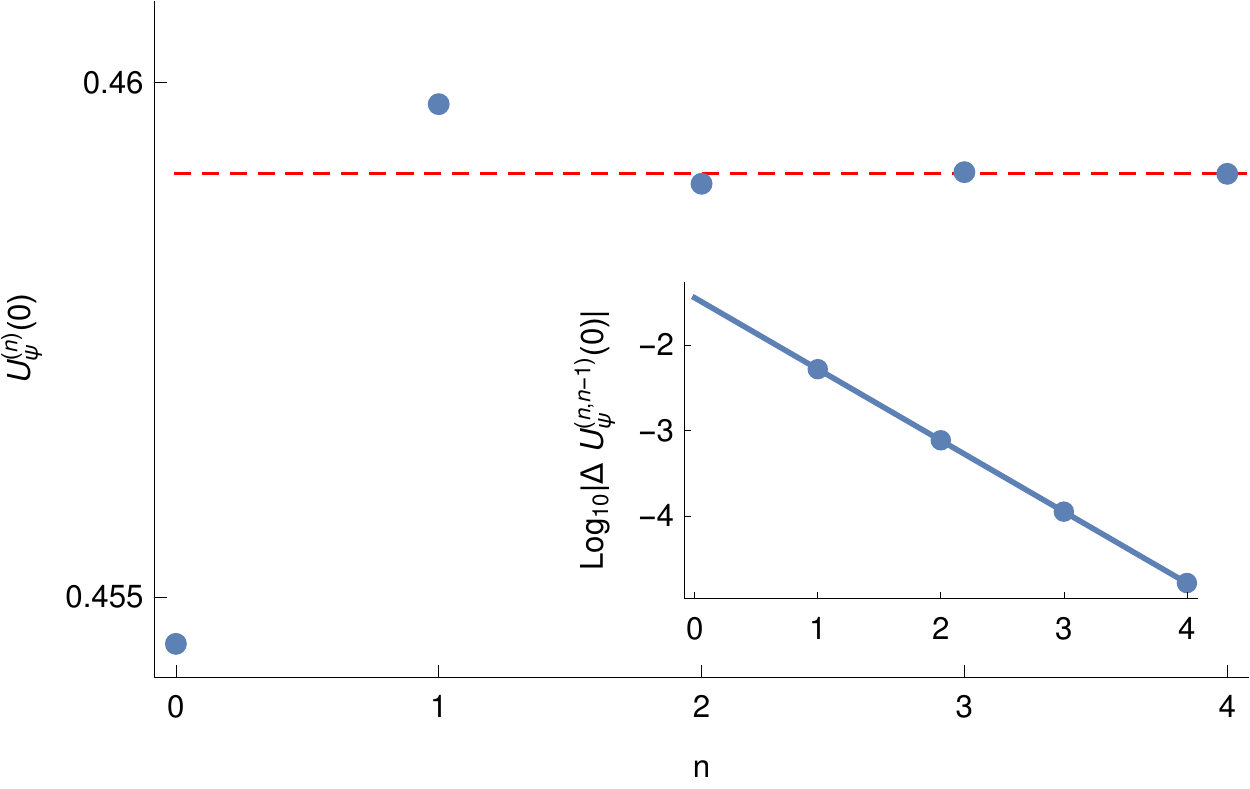}

\caption{Peak value of the approximation versus order $n$ for the nonlinear Camassa-Holm DE. The numerically exact peak value (red, dashed line) is also shown. \textbf{Inset:} A $\logten$ semi-log plot of the increments  $|\Delta U_\psi^{(n,n-1)}(0)|$ of the approximations, and a linear fit. Parameter values are $c = -3/2$, $A = 1/2$, and $K = 1/2$.}
\label{fig:ch_max_k05}
\end{figure}

The fast convergence to the correct answer is conspicuous by inspecting the peak values (at $z=0$) in Fig.\ref{fig:ch_zoom_k05}, as Fig. \ref{fig:ch_max_k05} shows. The increments  $|\Delta U_\psi^{(n,n-1)}(0)| \equiv |U_\psi^{(n)}(0) - U_\psi^{(n-1)}(0)|$ decay to zero exponentially rapidly, as is seen in the semi-log plot in the inset of Fig. \ref{fig:ch_max_k05}. In each iteration almost an order of magnitude is gained in precision. We verified that this holds uniformly for all $z$.

We have verified that the iteration procedure converges exponentially rapidly to the correct answer for values of $K$ in the range $10^{-3} \leq K < 10^3$. So we find that the method works well far beyond the scope envisaged in \cite{BLUES} and this suggests that the approach does not require perturbation theory.

Our second example pertains to the propagation and diffusion of disturbances in liquids, gases, ..., traffic, etc. 
We start from one of the most widely used linear DEs in physics, the equation describing the diffusion of a (generalized) density $u(x,t)$, being $u_t  - \nu \,u_{xx} = 0$,
with $\nu$ the diffusion coefficient. We are interested in traveling wave solutions for general source terms (representing, e.g., external forces) and calculate the Green function in the co-moving coordinate $z=x-ct$. After scaling the variables (setting $\nu = 1$) we solve the ODE
\begin{equation}
  \label{eq:burgers_linear}
\mathcal{L}_zU \equiv  - U_z - \frac{1}{c} U_{zz} = \delta(z),
\end{equation}
with wavefront boundary conditions  $U_z(z\rightarrow - \infty) = 0$ (and $U(z\rightarrow - \infty) > 0$) and $U(z\rightarrow\infty) = 0$.
The exact solution (in every point including $z=0$) is the piecewise analytic exponential tail,
\begin{equation}
    \label{eq:burgers_BLUES}
    B(z) = \begin{cases}
    1, & z<0 \\
    \mathrm{e}^{-z/k}, & z\geq0
    \end{cases}
\end{equation}
and the wavefront velocity is $c(k) =1/k$. 

Now we propose to include nonlinearity {\em of any kind}, as long as the solution satisfies the boundary conditions. For example, including the convective nonlinearity of the equations of fluid motion, we arrive at Burgers' equation,
\begin{equation}
u_t + u\,u_x - \nu \,u_{xx} = 0
\end{equation}
which has proven useful in a variety of problems in physics and continues to be the subject of intensive research \cite{Burgers}. Its extensions in the form of Euler and Navier-Stokes equations with source terms that represent mass, momentum and energy sources, are relevant, e.g., in nonlinear acoustics \cite{acous}. For traveling wavefronts in the co-moving frame, and with an arbitrary (normalized) source $\psi$, Burgers' equation can be rewritten as the nonlinear ODE
\begin{equation}
    \label{eq:burgers_rewrite}
    \mathcal{N}_zU \equiv - U_z + kU U_z - kU_{zz} = \psi(z),
\end{equation}
which is compatible with our boundary conditions provided $0<k<1/2$.

We now attempt to solve this nonlinear problem using superposition based on the Green function \eqref{eq:burgers_BLUES}.
Note that $B(z)$ now does {\em not} satisfy \eqref{conceptN} since the nonlinear term is not compensated. Notwithstanding this fact, we can still apply the iteration procedure because condition \eqref{conceptN} turns out not to be necessary. Indeed, \eqref{conceptNs} and \eqref{iterationstart} only require \eqref{conceptLs}. Therefore, we can still use $B(z)$ as BLUES function (although it does not satisfy the strict definition given in \cite{BLUES}). We investigate the convergence of the analytic approximations that result from iterating the residual operator, which now acts as $\mathcal{R}_z U = - k U U_z$. Obviously,  $\mathcal{R}_z B \neq 0$. Consequently, in the limit that  $\psi(z)$ approaches a Dirac delta source, the solution $B\ast\phi$  converges to a function different from, but isomorphic to $B(z)$,
\begin{equation}
B\ast \phi\, \sim \,B + B\ast  {\cal R}_z (B + B\ast {\cal R}_z (B+ B\ast {\cal R}_z (B + ... ))),
\end{equation}
which is straightforward to calculate. 

For an arbitary source $\psi$, the zeroth-order approximation is $B\ast\psi$. Choosing the exponential corner source \eqref{eq:source}, we obtain 
\begin{equation}
    \label{eq:burgers_conv}
    (B\ast\psi)(z) = \frac{1}{2} \begin{cases}
        2 - \frac{K}{K+k}\mathrm{e}^{z/K}, & z<0\\ 
        \frac{K}{K-k}\mathrm{e}^{-z/K} - \frac{2k^2}{K^2 -k^2}\mathrm{e}^{-z/k}, & z\geq0\\
    \end{cases}
\end{equation}
The calculation of higher-order order solutions is straightforward. 

\begin{figure}[!ht]
\begin{subfigure}{0.60\linewidth}
    \centering
    \includegraphics[width = \linewidth]{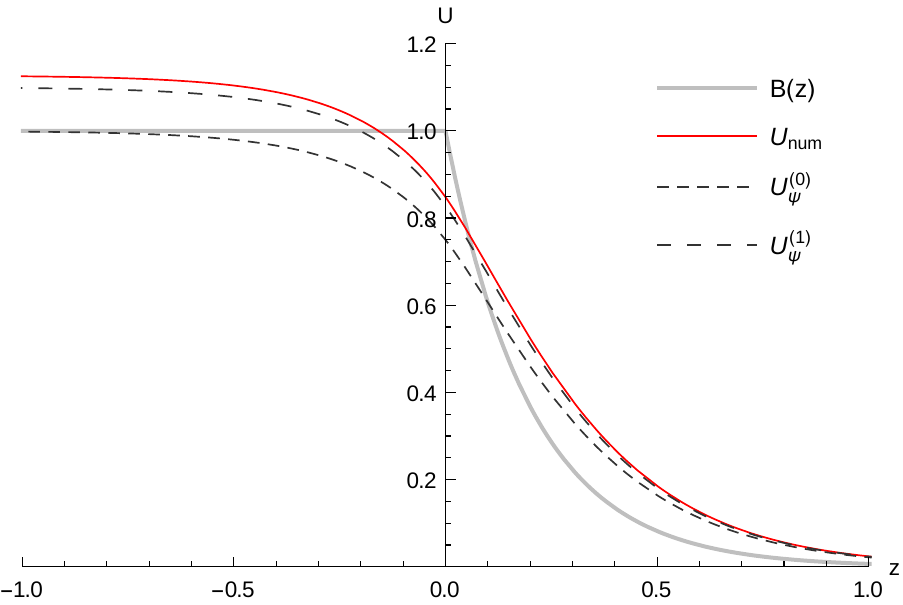}
    \caption{}
    \label{fig:B_k05}
\end{subfigure}
\begin{subfigure}{0.60\linewidth}
    \centering
    \includegraphics[width = \linewidth]{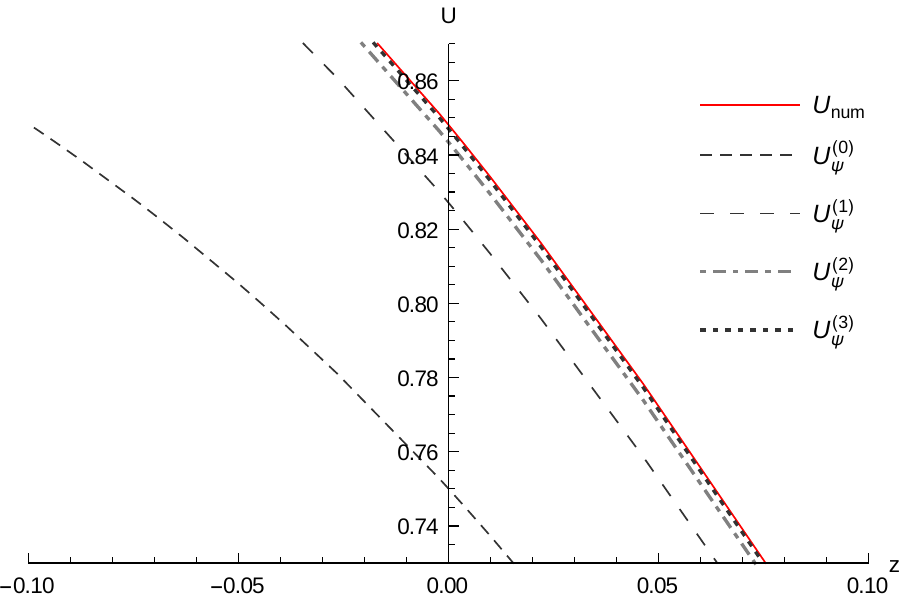}
    \caption{}
    \label{fig:B_zoom_k05}
\end{subfigure}
\caption{\textbf{(a)} Travelling wavefront solution to the nonlinear Burgers DE \eqref{eq:burgers_rewrite} with an exponential corner source \eqref{eq:source}. The numerical solution $U_{\rm num}$ (red, full line), the zeroth-order $U_\psi^{(0)}$ (black, dashed line) and the first-order approximation $U_\psi^{(1)}$ (black, wider spaced dashed line) are compared. The BLUES-function (gray) is also shown. \textbf{(b)} A zoomed-in view around the shoulder of the wavefront. The approximations are shown up to and including 3rd order. On this scale $U_\psi^{(3)}$ is practically on top of the numerical solution. Parameter values are $k = K = 1/5$.}
\end{figure}

In the following figures we compare the approximate analytical solutions with the numerically exact solution of the nonlinear DE  \eqref{eq:burgers_rewrite} with source \eqref{eq:source}. In Fig.\ref{fig:B_k05} the BLUES function, the zeroth and first-order approximations are shown, together with the numerical solution. In Fig.\ref{fig:B_zoom_k05} a zoom is presented near the shoulder of the wavefront at $z=0$ and solutions are shown up till 3rd order. The parameters are $k=K=1/5$. Note that in Fig.\ref{fig:B_k05} the simple convolution (zeroth-order) asymptotically approaches unity for $z \rightarrow - \infty$, as does the BLUES function. This differs from the exact asymptotic value, found by integrating the DE, $U_\psi(-\infty) = (1-\sqrt{1-2k})/k$ (which equals 1.1270... for $k=1/5$). The first-order approximation, $U_\psi^{(1)}(-\infty) = 1+ k/2$, is already a significant improvement.

\begin{figure}[!ht]
\begin{center}
    \includegraphics[width = 0.60\linewidth]{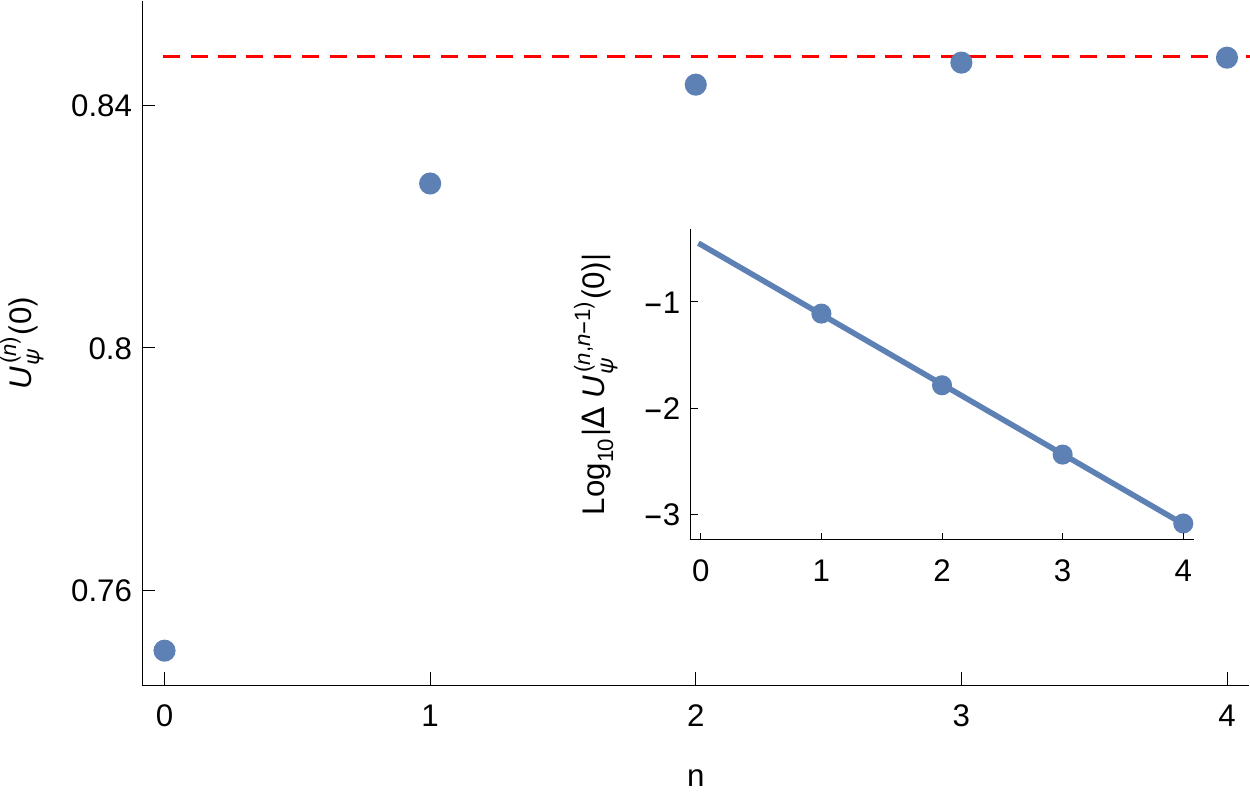}
   
\end{center}
\caption{Wavefront values $U_\psi^{(n)}(0)$ versus order $n$ for the nonlinear Burgers DE. The numerically exact value (red, dashed line) is also shown. \textbf{Inset:} A $\logten$ semi-log plot of the increments  $|\Delta U_\psi^{(n,n-1)}(0)|$ of the approximations, and a linear fit. Parameter values are $k = 1/5$ and $K = 1/5$.}
 \label{fig:B_max_k05}
\end{figure}

In order to assess the convergence we inspect the values near the shoulder of the wavefront (at $z=0$) in Fig.\ref{fig:B_zoom_k05}. The results are shown in Fig. \ref{fig:B_max_k05}. The convergence to the numerically exact value is fast and monotonic. The increments  $|\Delta U_\psi^{(n,n-1)}(0)| \equiv |U_\psi^{(n)}(0) - U_\psi^{(n-1)}(0)|$ decay to zero exponentially rapidly, as the semi-log plot in the inset of Fig. \ref{fig:B_max_k05} shows. In each iteration almost an order of magnitude is gained in precision.  Apparently, whether or not condition \eqref{conceptN} is satisfied does not affect the fast convergence of the method.

We verified that the iteration procedure converges exponentially rapidly to the correct answer for values of $K$ in the range $10^{-3} \leq K < 10^3$. Furthermore, for other types of nonlinearity which we tried, and for different choices of sources, the method also works (under the same boundary conditions).

In conclusion, an analytic iteration procedure based on the recently proposed BLUES function method holds promise to provide useful explicit solutions to a presumably large variety of nonlinear DEs of general interest for physicists. The class of equations which can be discussed with the method consists at this stage of ordinary DEs that are deterministic and integrable. Furthermore, the boundary conditions must require the solution, or its derivative, to decay to zero asymptotically (for large $|z|$). This conclusion has been reached along three lines of progress. 

Firstly, for a nonlinear ODE with a delta source that possesses an exact solution, which is at the same time a Green function for a related linear ODE, an iteration has been set up which converges exponentially rapidly to the correct solution for an arbitrary source. This has been demonstrated for the Camassa-Holm equation for traveling wave pulses (solitons) on shallow water. 

Secondly, and of general interest, is our observation that a Green function of a linear problem can be used as BLUES function in an analytic approach for solving a nonlinear ODE. Nonlinear terms can be added rather freely to the ODE as long as the solution respects the boundary conditions. For the nonlinear ODE with an arbitrary source an exponentially rapidly convergent sequence of analytic solutions can be calculated. The zeroth-order approximation is a simple convolution product and is already useful in practice. The first-order approximation is more interesting because it contains the effects induced by the nonlinear terms in the ODE. The first-order approximation can be calculated with moderate effort and provides already an accurate solution as compared with the numerically exact one. This has been demonstrated starting from the linear diffusion equation applied to traveling wavefronts and choosing Burgers' equation as the related nonlinear DE. We verified that the approach also works for other nonlinear extensions (not discussed here). The freedom of choice of the nonlinearity is a surprising and useful property which is likely to lead to applications in diverse domains. 

Thirdly, the iteration procedure developed here does not require the presence of a small parameter in which an expansion is performed. The examples discussed indicate that the BLUES function method does not require perturbation theory and can be used more generally.

In closing, we look out to the following applications in particular. In the field of nonlinear optics, laser frequency combs are used for time-keeping, metrology or spectroscopy. These combs can be generated by making use of the nonlinear Schr\"odinger equation (NLSE) with a source term \cite{Kippenbergeaan8083,Yang,ASL}. In semiconductor physics, solitons in a polariton condensate present a novel technique for information storage and communications. To study the dynamics of such a condensate, a dissipative Gross-Pitaevskii (GP) model with an external source (pump) can be used \cite{Xuekai2017,Zhenya}. When considering applications in the field of microfluidics, liquid films play a significant role. Therefore, it is important to study the stability and steady-state behaviour of these films under the influence of mechanical or thermal factors \cite{Jutley2018}, which can for instance be modeled by sources or sinks. When dealing with solitary waves on shallow water, calculating physically relevant profiles of tidal bores and exploring whether the BLUES function method can be used to refine the recent ``minimal analytic model" for this problem \cite{Berry}, is a challenge. In the theory of superconductivity, we envisage applying the method to the sine-Gordon equation, which is used to describe the physics of fluxons in long Josephson junctions under the influence of a driving force \cite{Gonzalez,Gul}. In the area of interface growth, the Kardar-Parisi-Zhang equation and the related linear stochastic heat equation come to mind as candidates for application \cite{Takeuchi}. One can think of more examples of nonlinear DEs with sources that can be investigated with the proposed method. One can also envisage a possible extension of the method to nonlinear partial DEs (with a first derivative with respect to time) and to stochastic (non-deterministic) DEs.

{}

\begin{thebibliography}{}
\bibitem{error} Our present formulation corrects an error in Eqs. 35 and 36 in the original formulation presented in \cite{BLUES}
\end{thebibliography}
\end{document}